\documentclass[12pt,twoside,dvips]{article}
\usepackage{amssymb}
\usepackage{pifont}
\usepackage{amsfonts}
\usepackage{amsmath}
\usepackage[spanish,english]{babel}
\usepackage{graphicx}

\begin{document}

\title{$Z^{\prime}$ boson signal at Tevatron
and LHC in a 331 model}
\author{N. Gutierrez, R. Mart\'{\i}nez$\thanks{%
e-mail: remartinezm@unal.edu.co}$ \ and F. Ochoa$\thanks{%
e-mail: faochoap@unal.edu.co}$ \and Departamento de F\'{\i}sica, Universidad
Nacional, \\
Bogot\'{a}-Colombia}
\maketitle

\begin{abstract}
We analyse the possibilities to detect a new $Z^{\prime}$ boson in di-electron events at Tevatron and LHC in the framework of the 331 model with right-handed neutrinos. Using $p\bar{p}$ collision data collected by the CDF II detector at Fermilab Tevatron, we find that the 331 $Z^{\prime}$ boson is excluded with masses below $920$ GeV. For an integrated  luminosity of $100 fb^{-1}$ at LHC, and considering a central value $M_{Z^{\prime}}=1500$ GeV, we obtain the invariant mass distribution in the process $pp\rightarrow Z^{\prime}\rightarrow e^{+}e^{-}$, where a huge peak, corresponding to 800 signal events, is found above the SM background. The number of di-electron events vary from 10000 to 1 in the mass range of $M_{Z^{\prime}}=1000-5000$ GeV.         
\end{abstract}

\section{Introduction}

In many extensions of the Standard Model (SM), new massive and neutral gauge
bosons, called $Z^{\prime }$, are predicted \cite{zprimas}. The detection of a $Z^{\prime}$
resonance have became in a matter of high priority in particle physics,
which could reveal many features about the underlying unified theory.
Indirect search for these neutral bosons have been carried out at LEP,
through mixing with the $Z$ boson \cite{LEP}. Direct $Z^{\prime}$ production
have been currently proved at the Tevatron \cite{CDF,TEVA}. The discovery potential
for $Z^{\prime}$ particles have been explored at the forthcoming Large
Hadron Collider (LHC) \cite{ATLAS} in the $M_{Z^{\prime}}\approx 1-5$ TeV range.
The search for this particle have also been explored in the planned
International Liner Collider (ILC) \cite{ILC}.

There are many theoretical models which predict a $Z^{\prime}$ mass in the
TeV range, where the most popular are the $E_{6}$ motivated models \cite{zprimas,E6},
the Left-Right Symetric Model (LRM) \cite{LR}, the $Z^{\prime}$ in Little Higgs
scenario \cite{LH} and the Sequential Standard Model (SSM), which has heavier couplings than those of the SM Z boson. Searching for $%
Z^{\prime}$ in the above models has been widely studied in the literature \cite{zprimas} and
applied at LEP2, Tevatron and LHC. On the other hand, the
models with gauge symmetry $SU(3)_{c}\otimes SU(3)_{L}\otimes U(1)_{X},$
also called 331 models \cite{ten,frampton,twelve}, arise as an interesting alternative with $Z^{\prime}$
boson and many well-established motivations. First of all, from the
cancellation of chiral anomalies \cite{anomalias} and asymptotic freedom in
QCD, the 3-3-1 models can explain why there are three fermion families.
Secondly, since the third family is treated under a different
representation, the large mass difference between the heaviest quark family
and the two lighter ones may be understood \cite{third-family}. Thirdly, the
models have a scalar content similar to the two Higgs doublet model (2HDM),
which allow to predict the quantization of electric charge and the vectorial
character of the electromagnetic interactions \cite%
{quantum-charge,vectorlike}. Also, these models contain a natural
Peccei-Quinn symmetry, necessary to solve the strong-CP problem \cite%
{PC,PC331}. Finally, the model introduces new types of matter relevant to
the next generations of colliders at the TeV energy scales, which do not
spoil the low energy limits at the electroweak scale.

Although the theoretical and phenomenological features associated to the $%
Z^{\prime}$ boson is widely described in the literature \cite{zprimas}-\cite{ILC}, there are few
studies at Tevatron and LHC colliders of the $Z^{\prime}$ boson in the
framework of the 331 models \cite{barreto}. In this work we search for $Z^{\prime}$ bosons
in dielectron events produced in $p\bar{p}$ and $pp$ collisions for Tevatron
and LHC colliders, respectively in the framework of the 331 model with right-handed
neutrinos, which we denote as the Foot-Long-Truan (FLT) model \cite{twelve}.

\section{The 331 spectrum}

The fermionic structure is shown in Tab. \ref{tab:espectro} where all
leptons transforms as $(\mathbf{3,X}_{\ell }^{L})$ and $(\mathbf{1,X}_{\ell
}^{R})$ under the $\left(SU(3)_{L},U(1)_{X}\right) $ sector, with $\mathbf{X}%
_{\ell }^{L}$ and $(\mathbf{X}_{\ell }^{R})$ the $U(1)_{X}$ generators
associated with the left- and right-handed leptons, respectively; while the
quarks transforms as $(\mathbf{3}^{\ast }\mathbf{,X}_{q_{m^{\ast }}}^{L})$, $%
(\mathbf{1,X}_{q_{m^{\ast }}}^{R})$ for the first two families, and $(%
\mathbf{3,X}_{q_{3}}^{L})$, $(\mathbf{1,X}_{q_{3}}^{R})$ for the third
family, each one with its $U(1)_{X}$ values for the left- and right-handed
quarks. The quantum numbers $\mathbf{X}_{\psi }$ for each representation are
given in the third column from Tab. \ref{tab:espectro}, where the electric
charge is defined by

\begin{equation}
Q=T_{3}+\beta T_{8}+XI,  \label{charge}
\end{equation}

\noindent with $T_{3} = 1/2$diag$(1,-1,0)$, $T_{8}=(1/2\sqrt{3})$diag$%
(1,1,-2)$. The most popular cases are when $\beta=-1/\sqrt{3}$ and $-\sqrt{3}$, where the first case
contains the Foot-Long-Truan model (FLT) \cite{twelve} and the second
contains the Pisano-Pleitez-Frampton model (PPF) \cite{ten,frampton}.

\begin{table}[tbp]
\begin{center}
\begin{equation*}
\begin{tabular}{c||c||c}
\hline\hline
$representation$ & $Q_{\psi }$ & $X_{\psi }$ \\ \hline\hline
$\ 
\begin{tabular}{c}
$q_{m^{\ast }L}=\left( 
\begin{array}{c}
d_{m^{\ast }} \\ 
-u_{m^{\ast }} \\ 
J_{m^{\ast }}%
\end{array}%
\right) _{L}\mathbf{3}^{\ast }$ \\ 
\\ 
\\ 
$d_{m^{\ast }R};$ $u_{m^{\ast }R};$ $J_{m^{\ast }R}:\mathbf{1}$%
\end{tabular}%
\ $ & 
\begin{tabular}{c}
$\left( 
\begin{array}{c}
-\frac{1}{3} \\ 
\frac{2}{3} \\ 
\frac{1}{6}+\frac{\sqrt{3}\beta }{2}%
\end{array}%
\right) $ \\ 
\\ 
$-\frac{1}{3};$ $\frac{2}{3};$ $\frac{1}{6}+\frac{\sqrt{3}}{2}\beta $%
\end{tabular}
& 
\begin{tabular}{c}
\\ 
$X_{q_{m^{\ast }}}^{L}=\frac{1}{6}+\frac{\beta }{2\sqrt{3}}$ \\ 
\\ 
\\ 
$X_{d_{m^{\ast }},u_{m^{\ast }},J_{m^{\ast }}}^{R}=-\frac{1}{3},\frac{2}{3},%
\frac{1}{6}+\frac{\sqrt{3}}{2}\beta $%
\end{tabular}
\\ \hline\hline
\begin{tabular}{c}
$q_{3L}=\left( 
\begin{array}{c}
u_{3} \\ 
d_{3} \\ 
J_{3}%
\end{array}%
\right) _{L}:\mathbf{3}$ \\ 
\\ 
$u_{3R};$ $d_{3R};$ $J_{3R}:\mathbf{1}$%
\end{tabular}
& 
\begin{tabular}{c}
$\left( 
\begin{array}{c}
\frac{2}{3} \\ 
-\frac{1}{3} \\ 
\frac{1}{6}-\frac{\sqrt{3}\beta }{2}%
\end{array}%
\right) $ \\ 
\\ 
$\frac{2}{3};$ $-\frac{1}{3};$ $\frac{1}{6}-\frac{\sqrt{3}\beta }{2}$%
\end{tabular}
& 
\begin{tabular}{c}
\\ 
$X_{q^{(3)}}^{L}=\frac{1}{6}-\frac{\beta }{2\sqrt{3}}$ \\ 
\\ 
\\ 
$X_{u_{3},d_{3},J_{3}}^{R}=\frac{2}{3},-\frac{1}{3},\frac{1}{6}-\frac{\sqrt{3%
}\beta }{2}$%
\end{tabular}
\\ \hline\hline
\begin{tabular}{c}
$\ell _{jL}=\left( 
\begin{array}{c}
\nu _{j} \\ 
e_{j} \\ 
E_{j}^{-Q_{1}}%
\end{array}%
\right) _{L}:\mathbf{3}$ \\ 
\\ 
$e_{jR};$ $E_{jR}^{-Q_{1}}$%
\end{tabular}
& 
\begin{tabular}{c}
$\left( 
\begin{array}{c}
0 \\ 
-1 \\ 
-\frac{1}{2}-\frac{\sqrt{3}\beta }{2}%
\end{array}%
\right) $ \\ 
\\ 
$-1;$ $-\frac{1}{2}-\frac{\sqrt{3}\beta }{2}$%
\end{tabular}
& 
\begin{tabular}{c}
\\ 
$X_{\ell _{j}}^{L}=-\frac{1}{2}-\frac{\beta }{2\sqrt{3}}$ \\ 
\\ 
\\ 
$X_{e_{j},E_{j}}^{R}=-1,$ $-\frac{1}{2}-\frac{\sqrt{3}\beta }{2}$%
\end{tabular}
\\ \hline\hline
\end{tabular}%
\end{equation*}%
\end{center}
\caption{\textit{Fermionic content for three generations}\textit{. We take} $m^{\ast
}=1,2$ \textit{and} $j=1,2,3$}
\label{tab:espectro}
\end{table}

For the scalar sector, we introduce the triplet field $\chi $ with vacuum
expectation value (VEV) $\left\langle \chi \right\rangle ^{T}=\left( 0,0,\nu
_{\chi }\right) $, which induces the masses to the third fermionic
components. In the second transition it is necessary to introduce two
triplets$\;\rho $ and $\eta $ with VEV $\left\langle \rho \right\rangle
^{T}=\left( 0,\nu _{\rho },0\right) $ and $\left\langle \eta \right\rangle
^{T}=\left( \nu _{\eta },0,0\right) $ in order to give masses to the quarks
of type up and down respectively \cite{331us}.

In the gauge boson spectrum associated with the group $SU(3)_{L}\otimes
U(1)_{X},$ we are just interested in the physical neutral sector that
corresponds to the photon, $Z$, and $Z^{\prime },$ which are written in
terms of the electroweak basis for any $\beta$ as 
\cite{beta-arbitrary}

\begin{eqnarray}
A_{\mu } &=&S_{W}W_{\mu }^{3}+C_{W}\left( \beta T_{W}W_{\mu }^{8}+\sqrt{%
1-\beta ^{2}T_{W}^{2}}B_{\mu }\right) ,  \notag \\
Z_{\mu } &=&C_{W}W_{\mu }^{3}-S_{W}\left( \beta T_{W}W_{\mu }^{8}+\sqrt{%
1-\beta ^{2}T_{W}^{2}}B_{\mu }\right) ,  \notag \\
Z_{\mu }^{\prime } &=&-\sqrt{1-\beta ^{2}T_{W}^{2}}W_{\mu }^{8}+\beta
T_{W}B_{\mu },  \label{neutral bosons}
\end{eqnarray}

\noindent where the Weinberg angle is defined as \cite{beta-arbitrary}

\begin{equation}
S_{W}=\sin \theta _{W}=\frac{g_{X}}{\sqrt{g_{L}^{2}+\left( 1+\beta
^{2}\right) g_{X}^{2}}}
\end{equation}

\noindent and $g_{L},$ $g_{X}$ correspond to the coupling constants of the
groups $SU(3)_{L}$ and $U(1)_{X}$, respectively. It is to note that the $Z$
and $Z^{\prime}$ bosons in Eq. (\ref{neutral bosons}) are not truly masss
eigenstates, but there is a $Z-Z^{\prime}$ mixing angle that rotate the
neutral sector to the physical $Z_{1}$ and $Z_{2}$ bosons . However,
the hadronic reactions are much less sensitive to the $Z-Z^{\prime}$ mixing
than lepton reactions \cite{zprimas}. Thus, the $Z-Z^{\prime}$ mixing can be neglected and
we identify the $Z$ and $Z^{\prime}$ bosons as the physical neutral bosons.

\section{Neutral Couplings and the Cross Section}

Using the fermionic content from Tab. \ref{tab:espectro}%
, we obtain the neutral coupling for the SM fermions \cite{beta-arbitrary}

\begin{equation}
\mathcal{L}_{D}^{NC}=\frac{g_{L}}{2C_{W}}\left[ \overline{f}\gamma _{\mu
}\left( g_{v}^{f}-g_{a}^{f}\gamma _{5}\right) fZ^{\mu }+\overline{f}\gamma
_{\mu }\left( \widetilde{g}_{v}^{f}-\widetilde{g}_{a}^{f}\gamma _{5}\right)
fZ^{\mu \prime }\right] ,  \label{L-neutro}
\end{equation}

\noindent where $f$ is $U=(u,c,t),$ $D=(d,s,b)$ for up- and down-type
quarks, respectively and $N=(\nu _{e},\nu _{\mu },\nu _{\tau }),$ $L=(e,\mu
,\tau )$ for neutrinos and charged leptons, respectively. The vector and
axial-vector couplings of the $Z$ boson are the same as the SM Z-couplings

\begin{eqnarray}
g_{v}^{U,N} &=&\frac{1}{2}-2Q_{U,N}S_{W}^{2},\qquad \qquad g_{a}^{U,N}=\frac{%
1}{2},  \notag \\
g_{v}^{D,L} &=&-\frac{1}{2}-2Q_{D,L}S_{W}^{2},\qquad \quad \;g_{a}^{D,L}=-%
\frac{1}{2},  \label{Z-couplings}
\end{eqnarray}

\noindent with $Q_{f}$ the electric charge of each fermion given by Tab. \ref%
{tab:espectro}; while the corresponding couplings to $Z^{\prime }$ are given
by \cite{z2-decay}

\begin{eqnarray}
\widetilde{g}_{v,a}^{U,D} &=&\frac{C_{W}}{2\sqrt{1-\beta ^{2}T_{W}^{2}}}%
\left[ \frac{1}{\sqrt{3}}\left( diag\left( 1,1,-1\right) +\frac{\beta
T_{W}^{2}}{\sqrt{3}}\right) \pm 2Q_{U,D}\beta T_{W}^{2}\right] ,  \notag \\
\widetilde{g}_{v,a}^{N,L} &=&\frac{C_{W}}{2\sqrt{1-\beta ^{2}T_{W}^{2}}}%
\left[ \frac{-1}{\sqrt{3}}-\beta T_{W}^{2}\pm 2Q_{N,L}\beta T_{W}^{2}\right]
,  \label{Zprima-couplings}
\end{eqnarray}

\noindent where the plus sign ($+$) is associated with the vector coupling $%
\widetilde{g}_{v},$ and the minus sign ($-$) with the axial coupling $%
\widetilde{g}_{a}.$ The above equations are written for $\beta =-1/\sqrt{3}$%
, which corresponds to the FLT model. On the other hand, the differential
cross section for the process $pp(p\bar{p})\longrightarrow Z^{\prime
}\longrightarrow f\bar{f}$ is given by \cite{zprimas}

\begin{equation}
\frac{d\sigma }{dMdydz}=\frac{K(M)}{48\pi M^{3}}\sum%
\limits_{q}P[B_{q}G_{q}^{+}(1+z^{2})+2C_{q}G_{q}^{-}z],  \label{difcross}
\end{equation}

where $M=M_{ff}$ is the invariant final state mass, $z=\cos \theta $ the
scattering angle between the initial quark and the final lepton in the $Z^{\prime}$ rest frame, $%
K(M)\simeq 1.3$ contains leading QED corrections and NLO QCD
corrections, $y=1/2\log [(E+p_{z})/(E-p_{z})]$ the rapidity, $E$ the total
energy, $p_{z}$ the longitudinal momentum, $P=s^{2}/[(s-M_{Z^{\prime
}}^{2})^{2}+M_{Z^{\prime }}^{2}\Gamma _{Z^{\prime }}^{2}],$ $\sqrt{s}$ the
collider CM energy, $M_{Z^{\prime }}$ and $\Gamma _{Z^{\prime }}$ the $%
Z^{\prime }$ mass and total width, respectively. The parameters $B_{q}=[(%
\widetilde{g}_{v}^{q})^{2}+(\widetilde{g}_{a}^{q})^{2}][(\widetilde{g}%
_{v}^{f})^{2}+(\widetilde{g}_{a}^{f})^{2}]$ and $C_{q}=4(\widetilde{g}%
_{v}^{q}\widetilde{g}_{a}^{q})(\widetilde{g}_{v}^{f}\widetilde{g}_{a}^{f})$
contains the couplings from Eq. (\ref{Zprima-couplings}) for the initial quarks $q$ and the
final fermions $f$, while the parameter $G_{q}^{\pm
}=x_{A}x_{B}[f_{q/A}(x_{A})f_{\overline{q}/B}(x_{B})\pm f_{q/B}(x_{B})f_{%
\overline{q}/A}(x_{A})]$ contains the Parton Distribution Functions (PDFs) $%
f(x),$ and the momentum fraction $x=Me^{\pm y}/\sqrt{z}.$ We can
consider the Narrow Width Approximation (NWA), where the relation $\Gamma
_{Z^{\prime }}^{2}/M _{Z^{\prime }}^{2}$ is very small, so that the
contribution to the cross section can be separated into the $Z^{\prime}$
production cross section $\sigma(pp(\bar{p})\rightarrow Z^{\prime})$ and the
fermion branching fraction of the $Z^{\prime}$ boson $Br(Z^{\prime}%
\rightarrow f\bar{f})$

\begin{equation}
\sigma(pp(\bar{p})\rightarrow f\bar{f})=\sigma(pp(\bar{p})\rightarrow
Z^{\prime})Br(Z^{\prime}\rightarrow f\bar{f}),  \label{NWA}
\end{equation}

From the analysis of reference \cite{z2-decay} we can estimate that $\Gamma _{Z^{\prime }}^{2}/M
_{Z^{\prime }}^{2}\approx 1\times 10^{-4}$. Thus, the NWA is an appropriate approximation in our calculations.

\section{$Z^{\prime}_{FLT}$ at Tevatron}

A recent report of a search for electron-positron events in the invariant
mass range $150-950$ GeV collected by the CDF II detector at the Fermilab
Tevatron \cite{TEVA} have excluded possible $Z^{\prime}$ particles for five different
models: the $Z^{\prime}_{\eta}$, $Z^{\prime}_{\chi}$, $Z^{\prime}_{\psi}$
and $Z^{\prime}_{I}$ bosons from the $E_{6}$ model, and the $Z^{\prime}_{SM}$
from the Sequential Standard Model (SSM). We use the data from reference \cite{TEVA} in
order to bound the $Z^{\prime}_{FLT}$ mass from the FLT 331 model.
Basically, the CDF II is an azimuthally and forward-backward symmetric
particle detector, where the most important features and kinematical cuts are \cite{TEVA}

\vspace{0.3cm}

\ding{109} $p\bar{p}$ collisions at C.M. energy $\sqrt{s}=1.96$ TeV,

\vspace{0.3cm}

\ding{109} Integrated luminusity $L=1.3 fb^{-1}$,

\vspace{0.3cm}

\ding{109} Central Calorimeter in the $\left| \eta \right| \leq 1.1$ range
and plug Calorimeters in the $1.2 \leq \left| \eta \right| \leq 3.6$ range,

\vspace{0.3cm}

\ding{109} Transverse energy cut $E_{T} \geq 25$ GeV.

\vspace{0.3cm}

\begin{figure}[t]
\centering \includegraphics[scale=0.56]{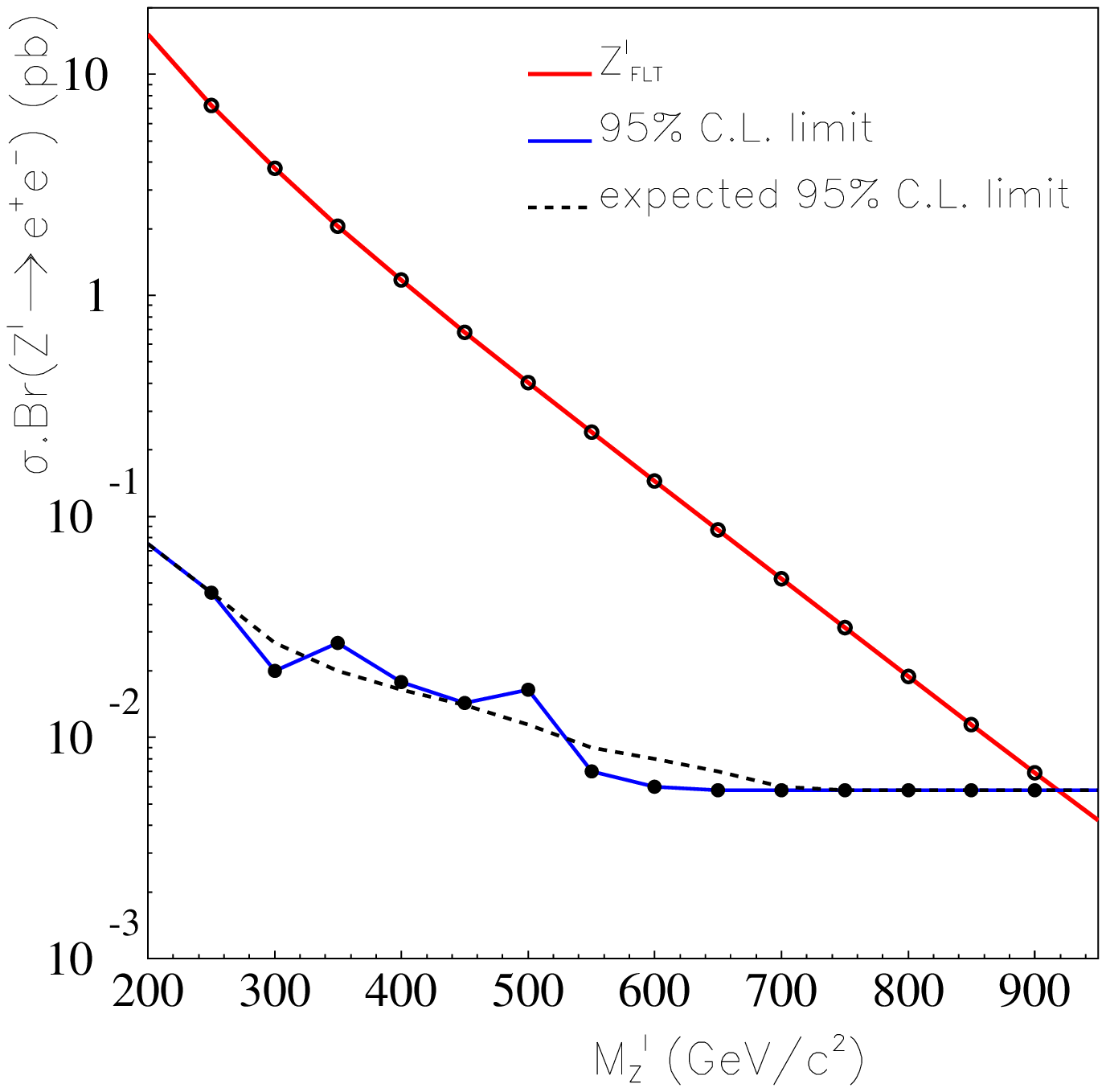}
\caption{\textsf{\ {\protect\small The limit on $\sigma Br$ as a function of the dielectron mass for the $95\%$ CL experimental data in Tevatron and the prediction of the 331 FLT model. Both plots cross at the bound $920$ GeV.}}}
\label{fig-teva}
\end{figure}

For this study, we use the CalcHep package \cite{calchep} in order to simulate $p\bar{%
p}\rightarrow e^{+}e^{-}$ events with the above kinematical criteria. Using a non-relativistic Breit-Wigner function and the CTEQ6M PDFs \cite{cteq}, we perform a numerical
calculation with the following parameters

\begin{equation}
\alpha ^{-1}=128.91,\quad S_{W}^{2}=0.223057,\quad \Gamma _{Z^{\prime
}}=0.02M_{Z^{\prime}},
\end{equation}

where the total width $\Gamma _{Z^{\prime }}=0.02M_{Z^{\prime }}$ is estimated from the
analysis performed in the reference \cite{z2-decay} for the FLT model. Fig. \ref{fig-teva}, shows the $%
95\%$ CL on $\sigma (pp(\bar{p})\rightarrow Z^{\prime })Br(Z^{\prime
}\rightarrow f\bar{f})$ extracted from reference \cite{TEVA} which does not show any
significant signal above the SM prediction. In the same plot, we show the
corresponding falling prediction for the $Z^{\prime }$ cross section in the
FLT model assuming that only SM fermions participate in the $Z^{\prime }$
decay. For small invariant masses, the 331 prediction exceeds the $95\%$ CL
bound. A bound is found at $M_{Z^{\prime }}=920$ GeV, where both
curves cross. Table \ref{tab:tevatron-bounds} shows lower bounds for different models at
Tevatron, where the 331 model exhibit a bigger bound than the $E_{6}$ $Z^{\prime }$ bosons.   

\begin{table}[tbp]
\begin{center}
\begin{tabular}{c||c|c|c|c|c|c}
\hline\hline
$Z^{\prime }-Model$ & $Z_{FLT}^{\prime }$ & $Z_{SM}^{\prime }$ & $Z_{\psi
}^{\prime }$ & $Z_{\eta }^{\prime }$ & $Z_{\chi }^{\prime }$ & $%
Z_{I}^{\prime }$ \\ \hline
\begin{tabular}{c}
$95\%C.L.$ $Bound$ \\ 
$(GeV)$%
\end{tabular}
& 920 & 923 & 822 & 891 & 822 & 729 \\ \hline\hline
\end{tabular}
\end{center}
\caption{\textit{$95\%$ CL lower bounds at Tevatron on the $Z^{\prime}$ mass for the FLT 331 model, the SSM, and $Z^{\prime}$ bosons from $E_{6}$ models}}
\label{tab:tevatron-bounds}
\end{table}

\section{$Z^{\prime}_{FLT}$ at LHC}

The design criteria of ATLAS at LHC could reveal $Z^{\prime }$ signal at the
TeV scale. The expected features of the detector are \cite{ATLAS}

\vspace{0.3cm}

\ding{109} $pp$ collisions at C.M. energy $\sqrt{s}=14$ TeV,

\vspace{0.3cm}

\ding{109} Integrated luminosity $L=100 fb^{-1}$,

\vspace{0.3cm}

\ding{109} Pseudorapidity below $\left| \eta \right| \leq 2.2$

\vspace{0.3cm}

\ding{109} Transverse energy cut $E_{T} \geq 20$ GeV.

\vspace{0.3cm}

\begin{figure}[t]
\centering \includegraphics[scale=0.52]{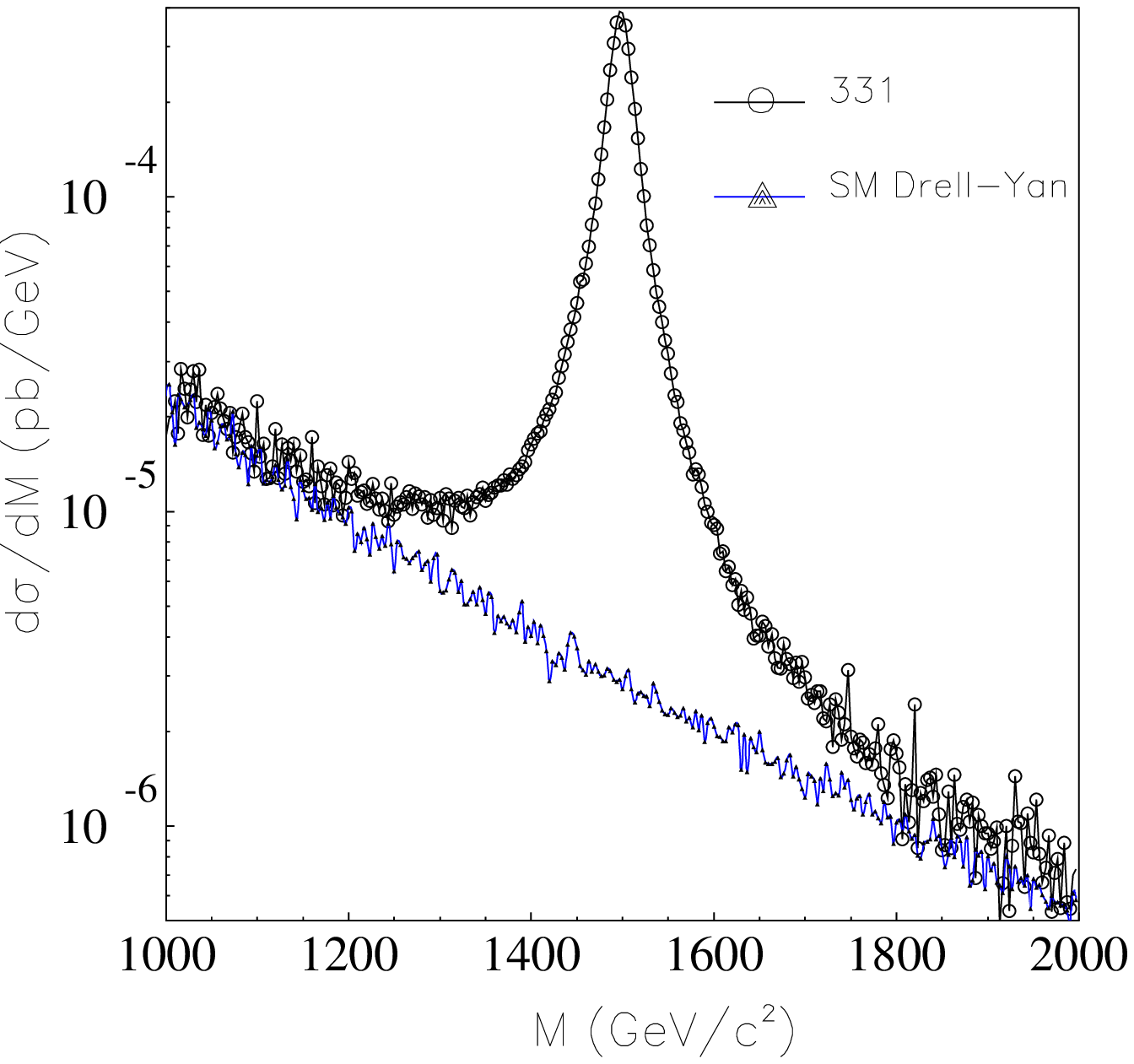} %
\includegraphics[scale=0.52]{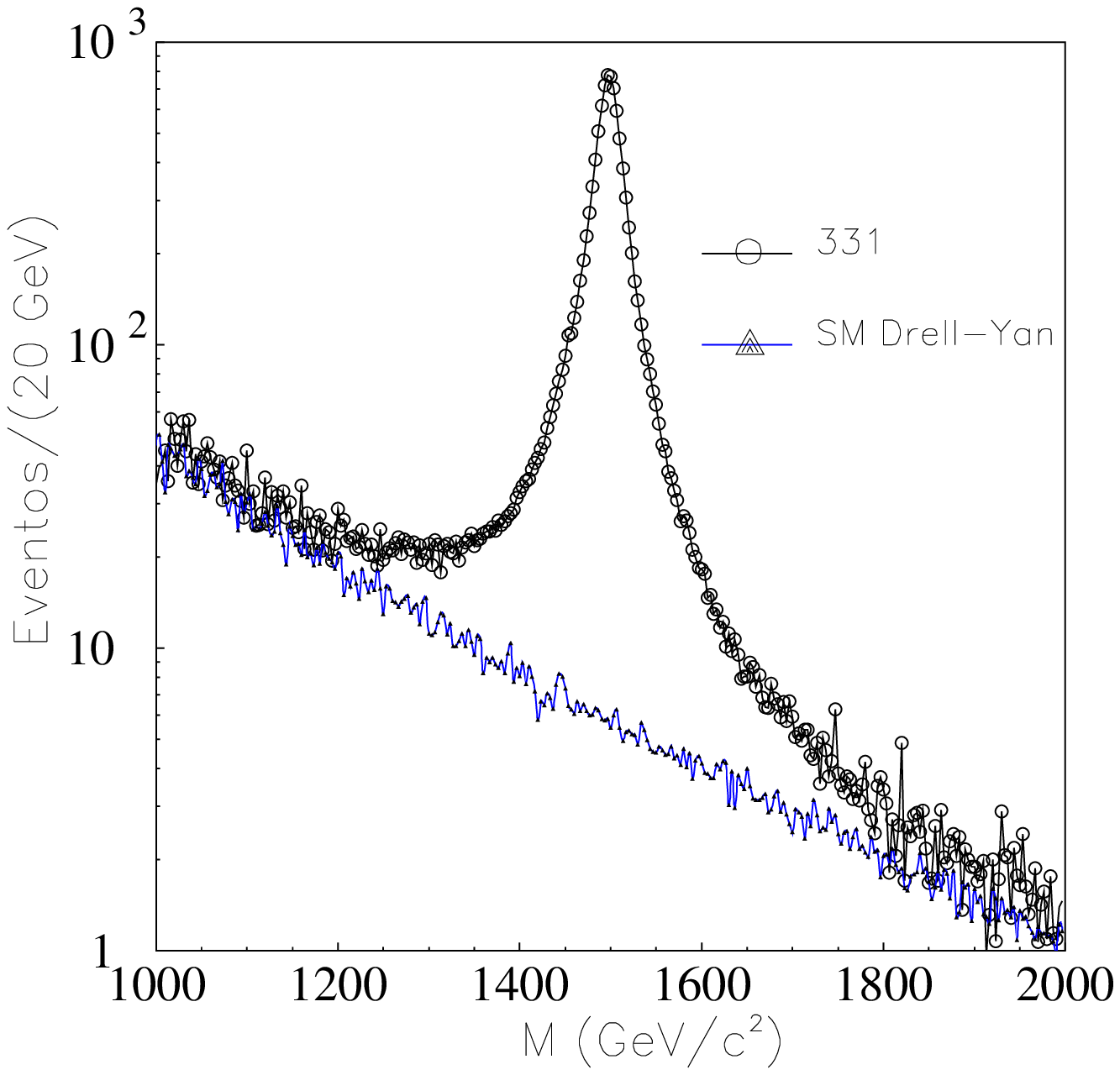}
\caption{\textsf{\ {\protect\small The left plot shows the cross section distribution as a function of the invariant final state mass for $M_{Z^{\prime}}=1500$ GeV in LHC. The right plot shows the number of events.}}}
\label{fig-dist}
\end{figure}

The left plot in Fig. \ref{fig-dist} shows the invariant mass distribution for the dielectron system as
final state, where we have chosen a central value $M_{Z^{\prime }}=1500$
GeV, which is a typical lower bound for FLT models from low energy analysis
at the $Z$-pole \cite{family-dependence}, and which falls into the expected detection
range for LHC. The right plot in the same figure shows the number of events for
the expected luminosity of $100 fb^{-1}$. We also calculate the SM Drell-Yan
spectrum in both plots with the same kinematical conditions, where we can see a huge peak
above the SM background in the resonance with about 800 signal events.

\begin{figure}[t]
\centering \includegraphics[scale=0.52]{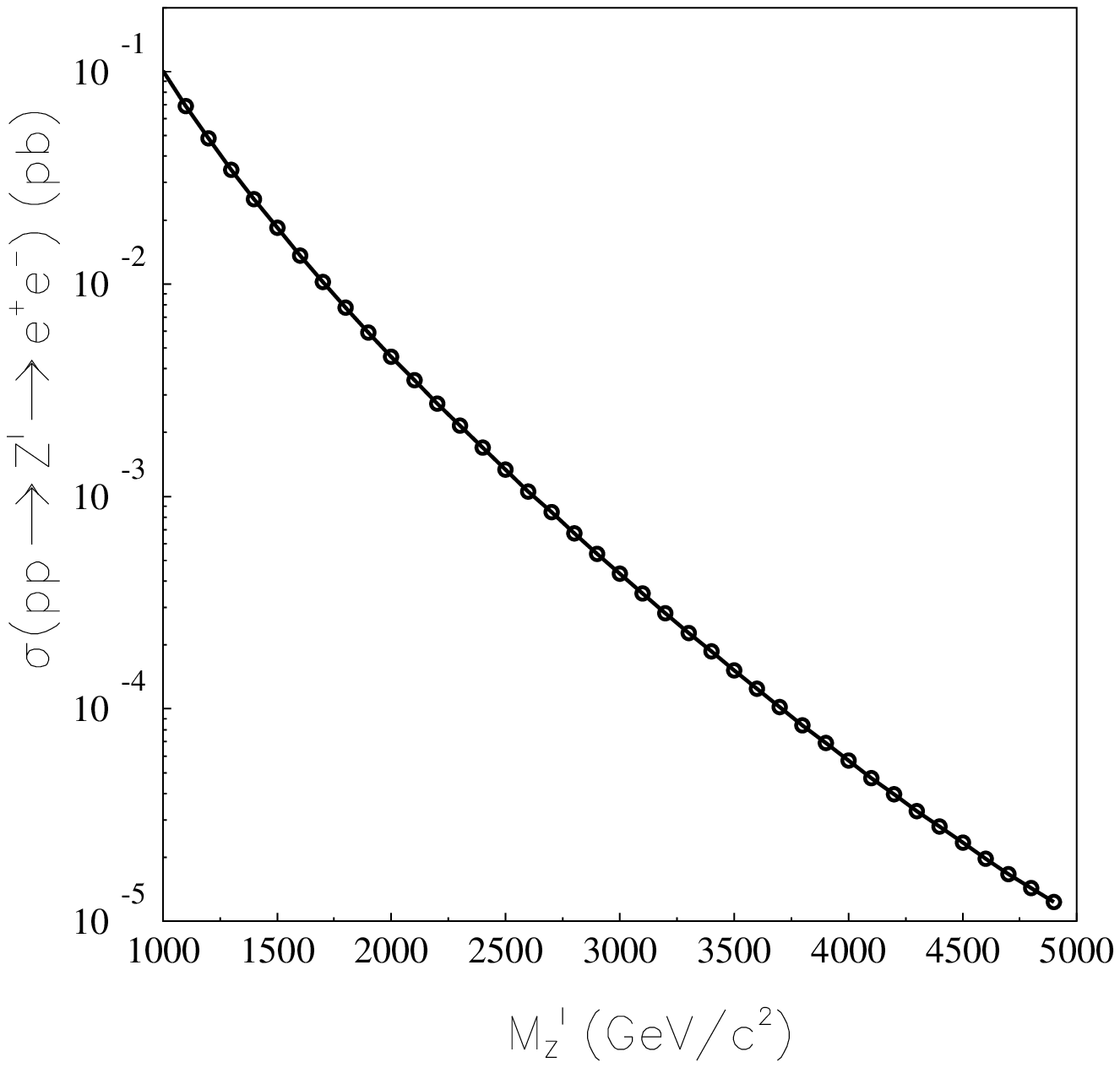} %
\includegraphics[scale=0.52]{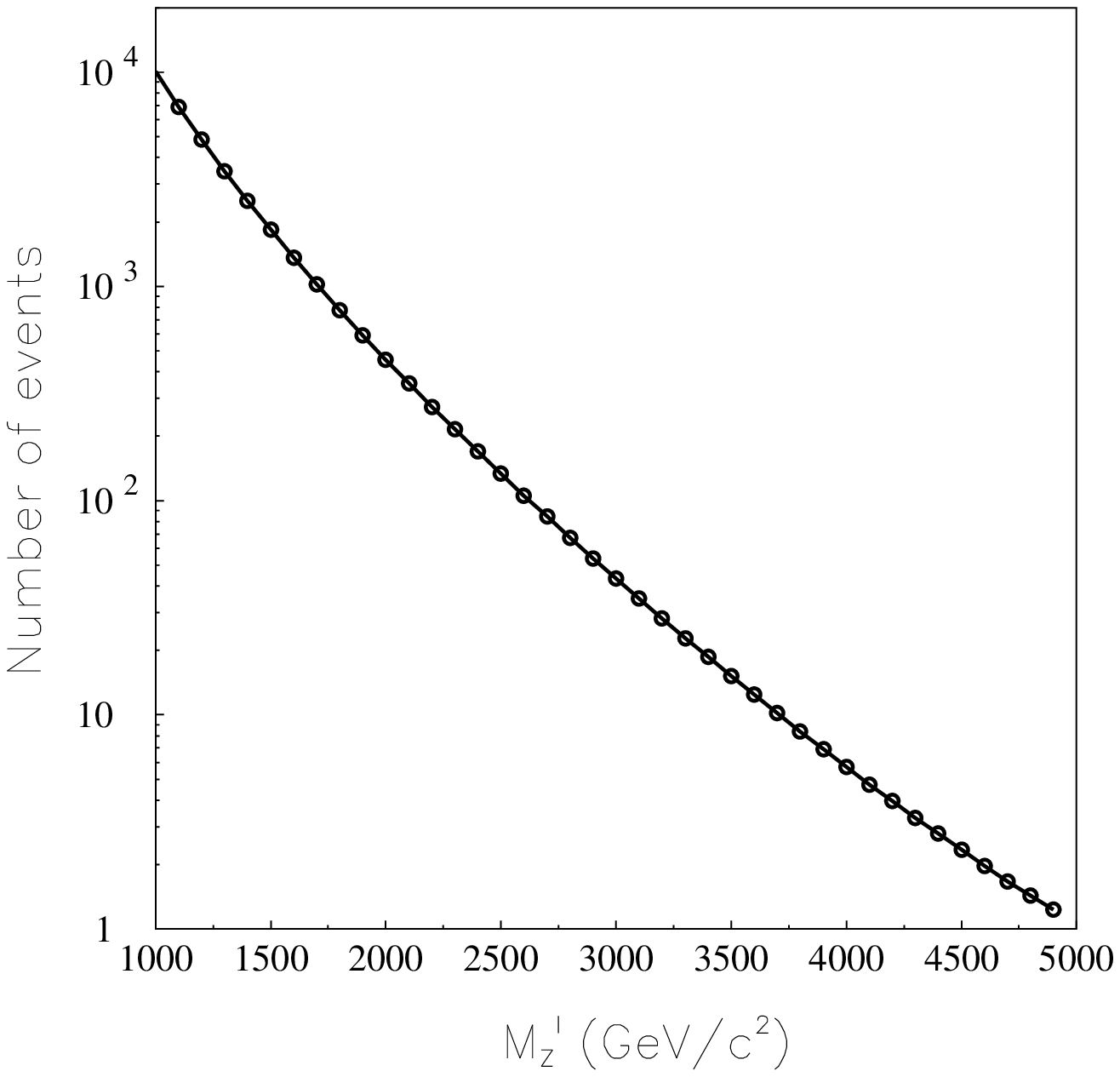}
\caption{\textsf{\ {\protect\small The left plot shows the cross section as a function of $M_{Z^{\prime}}$ in LHC. The right plot shows the number of events.}}}
\label{fig-cross}
\end{figure}

On the other hand, we calculate the cross section for the same leptonic channel as a function of $M_{Z^{\prime }}$, as shown in the left plot in Fig. \ref{fig-cross}. The right plot shows the number of
events, where the SM background is found to be essentially negligible for all the selected range. For $%
M_{Z^{\prime }}=1$ TeV, we get a huge number of events, corresponding to $%
10000$ signal events, while at the large mass limit $M_{Z^{\prime }}=5$ TeV,
we find just $1$ event per year.

\section{Conclusions}

In the framework of the FLT 331 model, we have analyzed the $Z^{\prime}$ production assuming the design criteria of CDF and ATLAS detectors at Tevatron and LHC colliders, respectively. Using recent data from CDF collaboration, the FLT $Z^{\prime}$ is excluded at $95\%$ CL with masses below $920$ GeV. For an integrated luminosity of $100 fb^{-1}$ in LHC and considering a central value of $M_{Z^{\prime}}=1500$ GeV, we find a narrow resonance with $800$ signal events above the SM background. If the $Z^{\prime}$ mass increases, the number of events decreases from $10000$ to $1$ signal event in the $M_{Z^{\prime}}=1000-5000$ GeV range.

\vspace{0.3cm}

This work was partially supported by Colciencias and by ALFA-EC funds through the HELEN programme.

\newpage

\end{document}